%% file: ms.tex
\documentclass[preprint2]{aastex}
\shorttitle{NGC 1333 IRAS 4A Outflow Variability}
\shortauthors{Choi et al.}
\received{}

\usepackage{times,epsfig}
\slugcomment{To appear in the Astrophysical Journal}
\newcommand{\HCOp}{{\rm HCO$^+$}}
\newcommand{\Ht}{{\rm H$_2$}}
\newcommand{\JJ}[2]{\mbox{$J = #1\rightarrow#2$}}
\newcommand{\kms}{\mbox{km s$^{-1}$}}
\newcommand{\ploteps}[2]{\centerline{\ifdim#2=0mm\epsfig{file=#1}%
                                     \else\epsfig{file=#1,width=#2}\fi}}
\newcounter{panel}[figure]
\renewcommand{\thepanel}
             {{\count20=1\advance\count20by\value{figure}%
               \stepcounter{panel}\small%
               \sc Fig. \the\count20\it\alph{panel}}}
\newcommand{\plotpanel}[3]
           {\vbox{\hsize=#2\epsfig{file=#1,width=#2}
                  \ifnum#3=1\makebox[\hsize]{\thepanel}\fi}}

\begin{document}
\fontsize{10}{10.6}\selectfont
\title{Variability of the NGC 1333 IRAS 4A Outflow:
       Molecular Hydrogen and Silicon Monoxide Images\altaffilmark{1}}
\author{\sc Minho Choi\altaffilmark{2,3},
            Klaus W. Hodapp\altaffilmark{4},
            Masahiko Hayashi\altaffilmark{5},
            Kentaro Motohara\altaffilmark{5,6},}
\author{\sc Soojong Pak\altaffilmark{2,7},
            and Tae-Soo Pyo\altaffilmark{5}}
\altaffiltext{1}{Based in part on data collected at Subaru Telescope,
                 which is operated
                 by the National Astronomical Observatory of Japan.}
\altaffiltext{2}{Korea Astronomy and Space Science Institute,
                 Hwaam 61-1, Yuseong, Daejeon 305-348, South Korea.}
\altaffiltext{3}{minho@kasi.re.kr.}
\altaffiltext{4}{Institute for Astronomy, University of Hawaii,
                 640 North A'ohoku Place, Hilo, HI 96720;
                 hodapp@ifa.hawaii.edu.}
\altaffiltext{5}{Subaru Telescope,
                 National Astronomical Observatory of Japan,
                 650 North A'ohoku Place, Hilo, HI 96720.}
\altaffiltext{6}{Institute of Astronomy, University of Tokyo,
                 Mitaka, Tokyo 181-0015, Japan.}
\altaffiltext{7}{Department of Astronomy and Space Science,
                 Kyung Hee University, Seocheon, Giheung,
                 Yongin, Gyeonggi 446-701, South Korea.}
\setcounter{footnote}{7}

\begin{abstract}
\fontsize{10}{10.6}\selectfont
The NGC 1333 region was observed in the H$_2$ 1--0 $S$(1) line.
The H$_2$ images cover a 5$'$ $\times$ 7$'$ region around IRAS~4.
Numerous H$_2$ emission features were detected.
The northeast-southwest bipolar outflow driven by IRAS 4A was studied
by combining the H$_2$ images with SiO maps published previously.
The SiO-H$_2$ outflows are continuous on the southwestern side
but show a gap on the northeastern side.
The southwestern outflow lobe curves smoothly,
and the position angle increases with the distance from the driving source.
The base and the outer tip of the northeastern outflow lobe
are located at positions
opposite to the corresponding parts of the southwestern lobe.
This point-symmetry suggests
that the outflow axis may be drifting or precessing clockwise
in the plane of the sky
and that the cause of the axis drift may be intrinsic to the outflow engine.
The axis drift model is supported
by the asymmetric lateral intensity profile of the SiO outflow.
The axis drift rate is $\sim$0\fdg011 yr$^{-1}$.
The middle part of the northeastern outflow
does not exactly follow the point symmetry
because of the superposition
of two different kinds of directional variability:
the axis drift of the driving source and the deflection by a dense core.
The axis drift model provides a good explanation
for the large deflection angle of the northeastern outflow.
Other H$_2$ emission features around the IRAS 4 region
are discussed briefly.
Some of them are newly found outflows,
and some are associated with outflows already known before.
\end{abstract}


\section{INTRODUCTION}

The NGC 1333 region contains
numerous young stellar objects and outflows
(Aspin et al. 1994; Hodapp \& Ladd 1995; Rodr\'{\i}guez et al. 1999).
It is one of the most active sites of star formation
in the solar neighborhood,
and the large number of active outflow sources
was described as a ``microburst'' of star formation (Bally et al. 1996).
Several protostars are located in the NGC 1333 molecular cloud,
and IRAS 4A is one of the brightest submillimeter sources among them
(Sandell \& Knee 2001).
IRAS 4A is a Class 0 protobinary system
(Sandell et al. 1991; Lay et al. 1995; Looney et al. 2000).

Single-dish observations showed
an interesting directional variability of the IRAS 4A molecular outflow
(Blake et al. 1995).
Detailed structure of the IRAS 4A molecular outflows
within $\sim$1$'$ of the driving source
was revealed by interferometric observations.
The interferometric maps showed
that the apparently large change of outflow direction
is partly owing to the existence of two outflows
driven by IRAS 4A (Choi 2001b)
and partly owing to the sharp bend (or deflection)
in the northeastern lobe of the main bipolar outflow
(Girart et al. 1999; Choi 2005a, hereafter Paper I).
Choi (2005a) suggested
that the sharp bend was caused by a collision
between the northeastern outflow
and a dense core in the ambient molecular cloud.
The jet-core collision model was supported by several lines of evidence
including the asymmetric morphology of the main outflow,
the HCN emission from shocked high-density gas near the impact point,
and the high-density core located in the course of the northeastern outflow
(Paper I; Choi 2005b).
However, the deflection cannot completely explain
the difference of position angles
between the SiO outflow near the driving source
and the large-scale ($>$ 1$'$) CO/\Ht\ outflow.

To investigate the overall structure
of the NGC 1333 IRAS 4A outflow system,
we observed the outflows in several molecular tracers.
In Paper I, we presented high angular resolution observations
in the SiO $v=0$ \JJ10\ line using the Very Large Array (VLA).
In this paper, we present our observations in the \Ht\ 1--0 $S$(1) line
using the University of Hawaii (UH) 2.2 m telescope
and the Subaru telescope.
In \S~2 we describe our \Ht\ observations.
In \S~3 we report the results of the \Ht\ imaging.
In \S~4 we compare the \Ht\ and the SiO images
and discuss the variability of the IRAS 4A main outflow.

\section{OBSERVATIONS}

\subsection{UH 2.2 {\rm m} Telescope Observations}

The molecular hydrogen emission features associated with IRAS 4
were included in a wide field image of the NGC 1333 region
in the \Ht\ 1--0 $S$(1) emission line
obtained in the nights of 1996 July 30 to August 3
using the QUIRC camera (Hodapp et al. 1996) at the UH 2.2 m telescope.
The individual integration time was 300 s.
The image was produced by stepping the telescope
in steps of 30$''$ in the north-south direction.
Therefore, most positions in the image were
covered by six individual frames for a total integration time of 30 min.
In the overlap region between adjacent vertical stripes,
the coverage is 12-fold for a total integration time of 60 min.

The UH \Ht\ image shows a set of artifacts
that are the result of the residual excess dark current
of the HgCdTe array used for the observations.
This particular array, one of the earliest HAWAII-I devices,
showed a strong excess of dark current
in pixels previously exposed to high signal levels,
with a decay time of the order of many minutes.
As a result of this effect,
combined with our method of collecting the individual images of this mosaic
in a north-to-south stepping pattern,
bright sources, most prominently SVS 13,
are associated with a long string of residual images south of them,
in 30$''$ spacing, and with slowly decaying intensity.
The faint extended nebulosity discussed in this paper
is not affected by this artifact.

\subsection{Subaru Telescope Observations}

Imaging was carried out on 2000 September 13
using the Cooled Infrared Spectrograph and Camera
for OH-Airglow Suppressor (CISCO),
which was equipped with a 1024 $\times$ 1024 HgCdTe detector
covering a field of view of 110$''$ $\times$ 110$''$
with a plate scale of 0\farcs11 pixel$^{-1}$ (Motohara et al. 2002).
We obtained narrow band images
with and without \Ht\ 1--0 $S$(1) line emission.
The \Ht\ 1--0  filter was centered on $\lambda$ = 2.120 $\mu$m
with a width of 0.020 $\mu$m (FWHM),
while the N215 filter for continuum emission
was centered on 2.147 $\mu$m with a width of 0.021 $\mu$m (FWHM).
Sixteen exposures with 60 s each were made for each of the filters
by stepping the telescope in steps of 10$''$--20$''$
along the east-west and/or north-south directions,
resulting the total on-source exposure time of 960 s
for the central 90$''$ $\times$ 90$''$ region.
Each frame of \Ht\ 1--0 and N215 filter images was flat-fielded
using sky-flat data and was subtracted by a corresponding sky frame,
which was made for each of the two filters with median filtering.
It was found that all the extended features in the \Ht\ 1--0 filter image
arise from the \Ht\ 1--0 $S$(1) line
by comparing it with the N215 filter image.
We thus use only the \Ht\ 1--0 filter image in this paper.

\subsection{Astrometric Calibration}

Accurate astrometry is essential
to comparing the \Ht\ image with the SiO image.
Since the adjacent continuum emission was not subtracted,
the \Ht\ images include both the line and the continuum emission.
In the UH \Ht\ image,
20 field stars listed in the 2MASS Point Source Catalogue%
\footnote{This publication makes use of data products
from the Two Micron All Sky Survey (2MASS),
which is a joint project of the University of Massachusetts
and the Infrared Processing
and Analysis Center/California Institute of Technology,
funded by the National Aeronautics and Space Administration
and the National Science Foundation.}
were identified.
In the Subaru \Ht\ image, only three 2MASS stars were found.

With the 2MASS coordinate data,
the {\it geomap} package of {\it IRAF} was used
to find the spatial transformation functions.
Then the {\it geotran} package was used to transform the observed images.
Comparing the stellar coordinates
of the transformed UH \Ht\ image and of the 2MASS catalogue,
the rms of position difference is 0\farcs19.
This value is about the same order of the position uncertainties
in the 2MASS catalogue,
which confirms that the transformation is acceptable.

Angular resolutions of the resulting images were measured
by fitting Gaussian intensity profiles to stellar objects.
The angular resolutions (FWHM) are
0\farcs82 for the UH \Ht\ image and 0\farcs46 for the Subaru \Ht\ image.

\subsection{VLA Data}

Details of the VLA observations and the results were presented in Paper I.
The NGC 1333 IRAS 4 region was observed using VLA
in the SiO $v=0$ \JJ10\ line.
The resulting images have a restoring beam of FWHM = 1\farcs96.
The SiO image covers a region up to $\sim$47$''$ from IRAS 4A.

\section{RESULTS}

Figure 1 shows the UH \Ht\ image of the NGC 1333 cloud
covering a 5\farcm2 $\times$ 7\farcm5 region.
The Subaru \Ht\ image (Fig. 2) has a better sensitivity
and a higher angular resolution than the UH image,
but the field of view is smaller.
In Figures 3$b$ and 4$a$, the VLA SiO map of the IRAS 4A outflows
is superposed on the \Ht\ images for comparison.

\input{fig1.tex} \input{fig2.tex}

\input{fig3.tex} \input{fig4.tex}

The structure of the \Ht\ emission features in the NGC 1333 region
was previously discussed by Hodapp \& Ladd (1995).
Recent observations in the submillimeter and centimeter continuum
(Rodr\'{\i}guez et al. 1999; Sandell \& Knee 2001)
provide helpful information for understanding the relation
between the \Ht\ emission features
and young stellar objects in this region.
The NGC 1333 region is crowded with outflows,
and the \Ht\ emission features should be inspected carefully
to avoid confusion.

Since the main interest of this paper
is the structure of the IRAS 4A molecular outflows,
we describe the \Ht\ emission features related to these flows first.
While the \Ht\ images show the large-scale structure of the main outflow,
little \Ht\ emission can be seen near the driving source,
probably owing to a high extinction through the dense core.
In contrast, the SiO line nicely shows
the outflow structure near the driving source,
probably owing to a high density.
Therefore, the \Ht\ images and the SiO map
provide complementary information.

On the northeastern side,
there is a relatively wide ($\sim$35$''$) gap
between ASR 57 and the northeastern end of the SiO outflow (Fig. 3$b$).
This gap is probably real
because CO outflow maps also show a similar gap
(see Fig. 3 of Blake et al. 1995).
A chain of \Ht\ features (ASR 57, HL 10, and HL 11 = HH 347A)
seems to belong to the IRAS 4A main outflow.
Each of these features is elongated in the general direction of the flow.
HH 347B is located away from the chain of \Ht\ features,
and it probably does not belong to the IRAS 4A outflow.

On the southwestern side,
the blueshifted SiO outflow is connected with the \Ht\ outflow,
and there is a considerable overlap (Fig. 4$a$).
Near the driving source,
the faint \Ht\ feature CHH 15 coincides with SiO outflow peak (SiOOP) 7.
The \Ht\ feature CHH 17 coincides with SiOOP 10 and 11
within a few arcseconds,
and the northeastern end of CHH 18 coincides with SiOOP 12 (Fig. 4$a$).
Further away from the driving source,
a chain of \Ht\ features (HL 5/3)
seems to belong to the IRAS 4A main outflow.

Along the southern SiO outflow,
the \Ht\ feature HL 6 is located in the middle of SiOOP 14, 15, and 16.
Unlike the main outflow,
the southern outflow shows no \Ht\ emission feature
beyond the SiO outflow lobe.
Therefore, the VLA SiO map probably traces
the whole extent of the southern outflow.

\subsection{Proper Motion}

Proper motions of \Ht\ features were measured
by comparing the Subaru and the UH images.
The coordinate systems of the two images were aligned
by comparing the positions of three stars near the \Ht\ features
(ASR 53, ASR 103, and CHH 18 IRS).
Since these stars are visible only in the near-IR or longer wavelengths,
they are either stars embedded in the NGC 1333 molecular cloud
or background stars,
and their proper motions may be negligible.
For bright \Ht\ features with well-defined peaks,
position offsets were measured by comparing pixels around the peaks,
typically pixels within $\sim$2$''$ from the peak positions.
Additional measurements were made by comparing the Subaru image
with the \Ht\ image of Hodapp \& Ladd (1995) in a similar manner,
which provides a longer time baseline but a larger uncertainty
owing to the worse angular resolution.
The two sets of measurements were consistent to each other,
and the average proper-motion vectors are shown in Figure 5.
The uncertainty of position offset was measured
from the residual position differences of the five stars
common to both images (the three stars above and ASR 54/55).

\input{fig5.tex}

The proper-motion vectors of HL 4 peaks
show a systematic motion toward southeast,
and those of HL 5 peaks show a motion toward southwest.
Assuming a distance of 320 pc (de Zeeuw et al. 1999),
the proper motion of HL 5 peaks ranges from 30 to 120 \kms\
with an uncertainty of 50 \kms.
In the discussions below (\S~4),
a proper motion of 100 \kms\ or 0\farcs07~yr$^{-1}$ is assumed
for calculations of the timescale of the IRAS 4A main outflow.

\section{DISCUSSION}

\subsection{Large-Scale Variability of the IRAS 4A Main Outflow}

As discussed in Paper I,
the northeastern SiO outflow shows a directional variability
that is probably caused by a collision with a dense core.
On top of this variability,
yet another directional variability is revealed
by combining the SiO map with the \Ht\ images.
This variability appears to affect
both the southwestern and the northeastern outflows
in a large ($\sim$5$'$) scale.
Since the northeastern outflow is more complicated
owing to the superposition of the two different variabilities,
we describe the southwestern outflow first.

The southwestern lobe of the SiO outflow appears very straight:
all the SiO outflow peaks are located
within a beam size from a straight line (see Fig. 2$a$ of Paper I).
When the SiO and the \Ht\ images are superposed, however,
the southwestern outflow appears to curve smoothly.
That is, the position angle (PA) of the outflow peaks
with respect to the driving source (IRAS 4A2)
increases with the angular distance (Fig. 3$a$ {\it top panel}).
Polynomial fits were drawn in Figure 3 to outline the flow.
A change of flow direction can be caused
either by intrinsic variability,
such as geometric change of the outflow engine,
or by external perturbation,
such as density gradients in the external medium.
The nearly constant gradient of the PA-distance relation suggests
that the large-scale variability is
probably owing to an intrinsic variability.
Another important clue to the variability mechanism
may come from the symmetry of the outflow morphology.

Near the driving source,
the direction of the northeastern outflow
is exactly opposite to that of the southwestern outflow.
The northeastern outflow changes direction abruptly
at a point $\sim$23$''$ away from the driving source,
and the position angle increases at a very high rate
(Fig. 3 {\it bottom panel}).
At a large distance, however, the PA-distance relation converges
back to the reverse of the southwestern flow.
Therefore, it appears that the southwestern and the northeastern outflows
would have shown a change of flow direction in a point-symmetric way,
if it were not for the jet-core collision of the northeastern outflow.
The symmetric morphology suggests
that the change of flow direction in the large scale
is caused by a geometric reconfiguration of the outflow engine.
That is, the axis of the IRAS 4A2 accretion disk may be changing direction,
and the outflow axis may be drifting%
\footnote{Outflows with changing directions
are often described as ``precessing''.
In this paper, we use the term ``drifting'' instead of precessing
because there is no evidence that the change is periodic.}
together.
The axis appears to drift clockwise in the plane of the sky.

The IRAS 4A outflow is not the only example
of point-symmetric, or {\sf S}-shaped, outflow
driven by young stellar objects.
Other examples include
the HH 340/343 outflow driven by NGC 1333 J032845.3+310542 (K)
and the \Ht\ jet driven by V380 Ori NE
(Hodapp et al. 2005; Davis et al. 2000).
Some outflows driven by high-mass (proto)stars
appear to have similar shapes,
e.g., IRAS 20126+4104 (Shepherd et al. 2000).

\subsection{Lateral Structure of the {\rm SiO} Outflow}

The lateral structure of the SiO outflow
was briefly described in \S~3.3 of Paper I.
Here we describe the structure in more detail
and discuss it in the context of the drifting axis.
Figure 4$b$ shows the lateral intensity profiles.
The SiO intensity profiles
of the undeflected part of the main outflow (peaks 4--12)
usually show a steep slope on the clockwise side
and a relatively slower slope (or shoulder) on the counterclockwise side.

The asymmetric slopes of the intensity profiles
can be explained by the drift of the flow axis.
Since the axis of the primary jet drifts clockwise,
the ambient molecular gas on the leading (clockwise) side
would be relatively fresh (and probably denser),
while the molecular gas on the opposite side
would have been already disturbed and accelerated in the past.
The shear zone between the primary jet and the ambient medium
may be thinner on the leading side than on the trailing side.
Then the SiO molecules in the shear zone of the leading side
would be more strongly shocked, excited more highly,
and make stronger emission,
which could produce the asymmetric intensity profile
across the outflow lobe.
On the trailing side,
the SiO emitting gas may be gradually cooling down
as the jet drifts away.
Therefore, the asymmetric lateral intensity profile provides
additional support for the drifting axis model.

\subsection{Drifting Axis of the IRAS 4A Main Outflow}

The PA-distance relation of the main outflow (Fig. 3$a$) suggests
that the amount of axis drift
within the observed extent of the main outflow
(i.e., from the vicinity of the driving source to HL 3)
is $\sim$20\arcdeg\
and that the rate of drift is $\sim$10\arcdeg\ arcminute$^{-1}$.
Assuming a proper motion of 0\farcs07 yr$^{-1}$ (see \S~3.1),
the temporal rate of axis drift is $\sim$0\fdg011 yr$^{-1}$.

Several authors suggested possible mechanisms
of intrinsic outflow variability (e.g., Eisl{\"o}ffel \& Mundt 1997;
Fendt \& Zinnecker 1998; Shepherd et al. 2000),
but most of them cannot generate a large amount of drift
as seen in the IRAS 4A outflow.
The tidal interaction between IRAS 4A1 and A2 cannot explain the drift
because the separation between them ($>$ 540 AU) is
too large to cause a strong effect.
In principle, A2 itself could be a close binary system
that has not been observationally resolved yet.
Assuming a proper motion of 0\farcs07 yr$^{-1}$,
the 130$''$ separation between IRAS 4A and HL 3b
gives an outflow timescale of $\sim$2000 yr.
Since the flow is not showing multiple turns,
the precession period would be longer than 4000 yr,
but it should be shorter than $\sim$16000 yr
to keep the precession angle reasonably small, say less than 90\arcdeg.
Since the precession period would be longer
than the binary orbital period by a factor of $\sim$20 (Bate et al. 2000),
the expected orbital period of the unresolved binary system is 200--800 yr.
The corresponding binary separation would be 30--80 AU,
assuming that the mass of IRAS 4A2 is $\sim$0.9 $M_\odot$,
which is about half the mass of IRAS 4A
(mass from Choi 2001b, scaled to 320 pc).
Then the angular separation would be 0\farcs09--0\farcs25,
which is too small for existing millimeter interferometers to resolve.
The accretion disk in such a system, however,
would probably be too small to drive an energetic outflow.

Another possible mechanism is the anisotropic accretion
that can occur
when the angular momentum of currently accreting mass is not parallel
to the rotation axis of previously accreted matter.
Considering that IRAS 4A is a binary system,
we propose the following scenario.
The initial IRAS 4A core had fragmented into two subcores (A1 and A2),
and each of them collapsed to form a protostar
and started an outflow activity.
While the total angular momentum of the A1/A2 system
may be parallel to the rotation axis of the initial core,
the spin axis of each individual protostar
(and the initial direction of each outflow)
could have been oriented differently.
(This scenario is consistent with the observations
because there is a considerable difference
between the directions of the main outflow driven by A2
and the southern outflow driven by A1.)
As the protostar accumulates mass from the large outer envelope,
the axis of the accretion disk would shift gradually
and converge to the direction of the average angular momentum
of the whole protostellar core.
This scenario can be tested in principle
by observationally determining the rotation axis of the IRAS 4A core,
but the velocity structure of this region
may be too complicated (Choi et al. 2004) for such a test.
In the anisotropic accretion model,
the change of direction may happen only once,
i.e., may not be periodic or oscillating,
which can explain
why most of the direction-changing outflows observed so far
do not show multiple turns.

Yet another possibility is the misalignment
between the rotation and magnetic fields.
Numerical simulations show
that the rotation axis, magnetic fields, and disk orientation
can precess significantly (Matsumoto \& Tomisaka 2004).
In such systems, changes of outflow direction may be observable.
The evolution of the field direction with respect to the rotation axis
may depend on the relative strength
between the rotation and the magnetic fields (Machida et al. 2006).

\subsection{Deflection of the Northeastern Outflow}

Theoretical models of jet-cloud collision predict
that, in the simple cases of steady jet,
the deflection process finishes in a few hundred years
either because the jet penetrates the dense clump
or because the jet is pinched off
(Raga \& Cant{\'o} 1995; de Gouveia Dal Pino 1999).
When the whole extent of the \Ht\ outflow is considered,
the timescale of the IRAS 4A main outflow is $\sim$2000 yr.
Then the deflection process of a steady jet
would have been almost finished in this timescale,
and the deflection angle would have been small.
The large deflection angle observed in the northeastern outflow
suggests that the real situation is more complicated than the simple models.
It was suggested that the deflection timescale can be longer
if the jet beam is not steady
so that the impact point does not stay at one position (Raga et al. 2002).
Therefore, the drifting axis model provides a good explanation
for the large deflection angle currently observed in the IRAS 4A outflow.

Assuming that the width of the (undeflected) primary jet at the impact point
is smaller than 4\farcs3 (deconvolved width of SiOOP 4),
and considering the axis drift rate in the previous section,
it takes less than $\sim$1000 yr
for the impact point at the ``surface'' of the dense core
to move across a jet width.
Since this crossing time is
comparable to or shorter than the cloud penetration timescale,
the large deflection angle can be maintained.

The deflection history of the northeastern outflow
can be summarized in the following scenario.
Choi (2005b) suggested
that the dense core obstructing the northeastern outflow is
located $\sim$50$''$ from IRAS 4A2
with a position angle of $\sim$16\arcdeg\ (Figs. 3$b$ and 6).
The detection of \HCOp\ and HCN emission suggests
that the obstructing core is very dense ($\sim$10$^5$ cm$^{-3}$)
(Paper I; Choi 2005b).
In the distant past,
the direction of the outflow had a large position angle,
and the flow would have passed by the eastern side of the obstructing core
with a large impact parameter
and would have been deflected only slightly.
For example, HL 11 would have had an (projected) impact parameter
of $\sim$19$''$ (or 6000 AU at 320 pc).
As the outflow axis drifts clockwise,
the impact point would move
on the ``surface'' of the dense core to the west,
and the impact parameter decreases.
Consequently, the outflow would be deflected more and more severely,
and the deflection angle may increase with time.
The position angle of the outflow axis
in the vicinity of the driving source is $\sim$19\arcdeg,
and the jet currently colliding against the dense core
may have an impact parameter of $\sim$3$''$ or 900 AU.
That is, the northeastern outflow is
currently making a nearly head-on collision against the obstructing core,
and consequently the deflection angle is large.

\input{fig6.tex}

Reipurth et al. (1996) suggested
that the HH 270/110 system may be an example of deflected outflow.
This interpretation, however, was questioned
because high-density molecular gas was not detected
near the proposed impact point (Choi 2001a).
Raga et al. (2002) suggested
that the apparent conflict can be reconciled
if the impact region is located
far away from the crossing point of the HH 270 and the HH 110 axes.
If the HH 110 flow is indeed a deflected segment of the HH 270 flow,
its deflection history must be more complicated than the case of IRAS 4A.
More detailed and sensitive observations are needed
to understand the HH 270/110 system better.

\subsection{Other \Ht\ Emission Features}

The \Ht\ images (Figs. 1 and 2) show several emission features
that were previously undetected or unlabeled.
Coordinates of their peak positions are listed in Table 1.
These \Ht\ features are labeled with a prefix CHH and briefly described here.
Some of the previously known \Ht\ features are also discussed below.

\input{tab1.tex}

{\it CHH 1.}---%
This is a short jet-like \Ht\ feature.
A probable driving source is VLA 14.

{\it CHH 2/4.}---%
These \Ht\ features and HH 347B probably belong to a single outflow.

{\it CHH 3.}---%
This is a short jet-like \Ht\ feature.
A probable driving source is ASR 74, the star just north of CHH 3.

{\it CHH 5.}---%
This is a linear chain of \Ht\ emission features.
Hodapp \& Ladd (1995) suggested
that CHH 5 and HL 4 are related with the SVS 13 complex.
Another possibility is
that the CHH~5--HL 4 flow may be driven by VLA 19 (SK 14).
Yet another possibility is that CHH 5b,
which is elongated roughly in the northeast-southwest direction,
may be a short outflow driven by VLA 19,
while CHH 5a/c (and HL 4) may belong to a separate outflow.

{\it CHH 6.}---%
This is an unusually straight jet-like \Ht\ feature.
It appears to be related with ASR 20/21 and probably with HH 344.

{\it CHH 7/10/14.}---%
These are isolated \Ht\ features with fuzzy appearances.
Their nature is unclear.
CHH 10 is associated with HH 759 (Walawender et al. 2005).

{\it CHH 8/9.}---%
These \Ht\ features seem to belong to the HH 5 system.

{\it CHH 11.}---%
This is a system of weak \Ht\ features.
It appears to be related with HL 8.

{\it CHH 12/13.}---%
These \Ht\ features probably belong to a single outflow.
A probable driving source is SK 1.

{\it CHH 16.}---%
This is a faint \Ht\ feature associated with SiOOP 17.
Since SiOOP 17 is kinematically distinct from the IRAS 4A main outflow,
it is not clear if it is related with the IRAS 4A system.
See \S~3.1 of Paper I for details.

{\it CHH 19.}---%
This \Ht\ feature probably belongs to the IRAS 4A outflow.

{\it HL 8.}---%
Hodapp \& Ladd (1995) suggested that HL 8 may be an extension of HL 4.
Considering the locations of SK1 and CHH 12/13, however,
HL 8 could be the counterpart of the CHH 12--13 outflow.

{\it HL 9.}---%
At $\sim$8$''$ south of IRAS 4BI,
the two peaks of HL 9 line up in the north-south direction,
which agrees with the direction of the IRAS 4BI bipolar outflow
seen in the HCN line (Choi 2001b).

{\it ASR 57.}---%
This is one of the brightest \Ht\ emission features in the region imaged.
The location of ASR 57 at the intersection
of the IRAS 4A outflow and the HH 7--11--HL 7 outflow
raised the possibility
that the two flows might be interacting (Hodapp \& Ladd 1995).
Since the dense core obstructing the IRAS 4A outflow
is a part of the large cloud
associated with SVS 13 and the HH 7--11 complex (Paper I; Choi 2005b),
the two outflows may be close to each other along the line of sight.
Further studies are needed to test
whether or not they are physically interacting indeed.
Proper motion studies of the \Ht\ peaks within ASR 57 may be useful.

{\it VLA 13/30/32.}---%
These centimeter continuum sources were detected
in the UH image as point-like sources.
The agreement between the VLA position and the near-IR position
is good within 0\farcs2.

\acknowledgements

We thank Mike Nassir for help with the reduction
of the near-infrared image from the UH 2.2 m telescope.
This work was partially supported by the LRG Program of KASI,
by the Korea-Japan Basic Scientific Cooperation Program of KOSEF,
and by the Japan-Korea Basic Scientific Cooperation Program of JSPS.

\vfill\centerline{\small\tt arXiv version}\vspace*{-\baselineskip}

\end{document}

%% file: fig1.tex
\begin{figure*}[p]
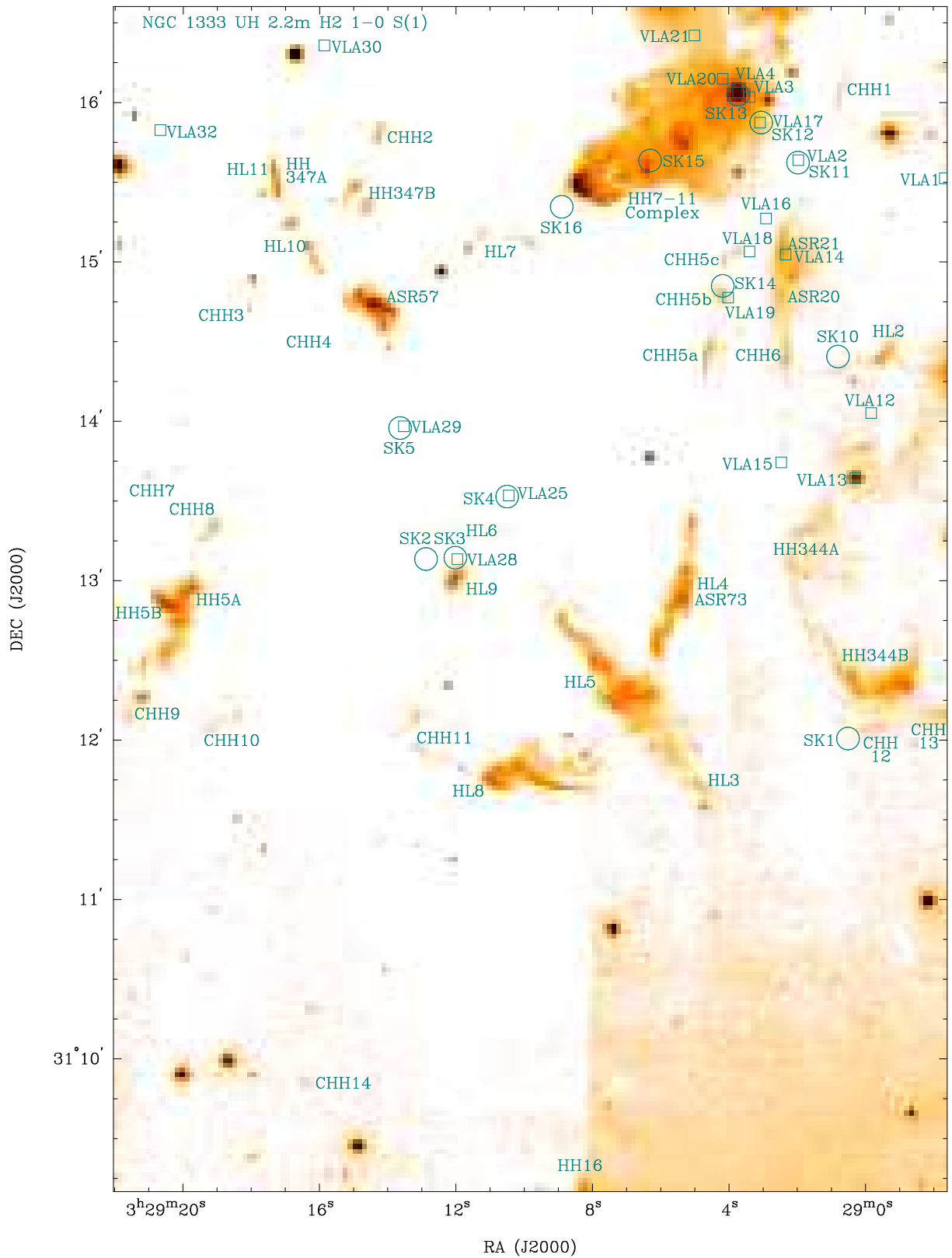

\ploteps{h2uh-lowq.eps}{161mm}
\caption{\small\baselineskip=0.825\baselineskip
Map of the \Ht\ 1--0 $S$(1) line toward the NGC 1333 IRAS 4 region
from the UH 2.2 m telescope.
Extended sources are labeled.
HL source numbers are from Hodapp \& Ladd (1995).
ASR source numbers are from Aspin et al. (1994),
and Herbig-Haro (HH) object numbers are
from Bally et al. (1996) and references therein.
Newly identified sources are numbered with a prefix CHH.
{\it Squares}:
Centimeter continuum sources
labeled following Table 1 of Rodr\'{\i}guez et al. (1999).
VLA 25 is associated with IRAS 4A,
and VLA 4 is associated with SVS 13.
{\it Open circles}:
Submillimeter continuum sources
labeled following Table 1 of Sandell \& Knee (2001).}
{\footnotesize
A preprint containing high-quality images of Figs. 1, 2, 3$b$, and 4$a$
can be found at {\verb+http://hanul.kasi.re.kr/~minho/Publications.html+}.}
\end{figure*}

%% file: fig2.tex
\begin{figure*}[!t]
\ploteps{h2subaru-lowq.eps}{185mm}
\caption{\small\baselineskip=0.825\baselineskip
Map of the \Ht\ 1--0 $S$(1) line toward the NGC 1333 IRAS 4 region
from the Subaru telescope.
Extended sources are labeled.
{\it Filled circles}:
Millimeter continuum sources in the IRAS 4 region (Looney et al. 2000).}
\end{figure*}

%% file: fig3.tex
\begin{figure*}[t]
\centerline{\plotpanel{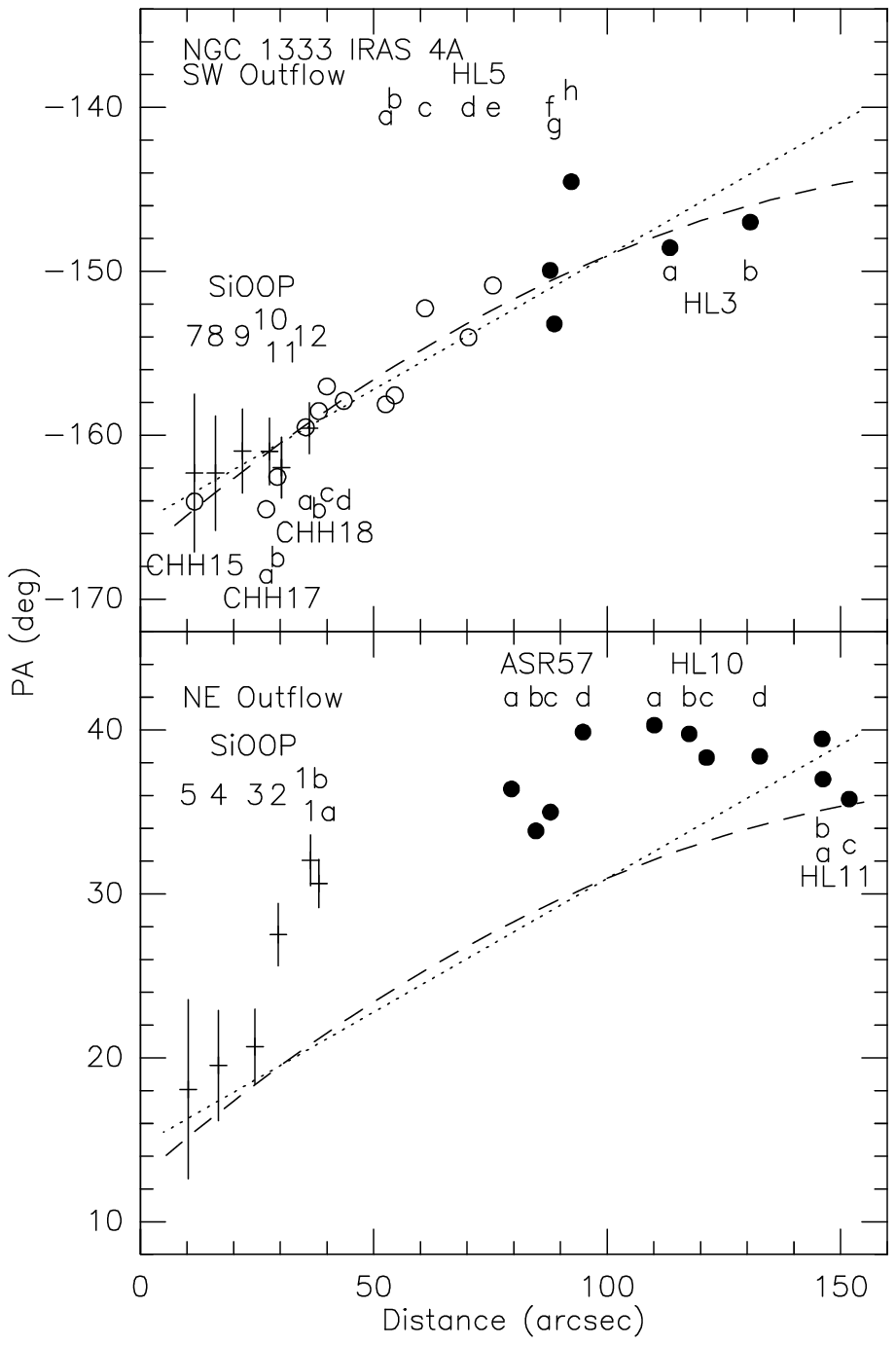}{78mm}{1}\hspace*{5mm}%
            \plotpanel{h2uh.sio-lowq.eps}{102mm}{1}}
\caption{\small\baselineskip=0.825\baselineskip
($a$)
Position angle of the emission peaks in the IRAS 4A main outflow
plotted as a function of the angular distance
with respect to the driving source (IRAS 4A2).
{\it Top panel}:
Southwestern outflow peaks.
{\it Bottom panel}:
Northeastern outflow peaks.
{\it Crosses}:
SiO outflow peaks (Paper~I; see Fig. 4).
The vertical size corresponds to the beam (FWHM).
{\it Open circles}:
\Ht\ emission peaks identified in the Subaru image.
{\it Filled circles}:
\Ht\ emission peaks identified in the UH image.
{\it Dotted straight line}:
Linear least-squares fit to the southwestern outflow peaks.
{\it Dashed curve}:
Parabolic least-squares fit to the southwestern outflow peaks.
In the bottom panel, the fits in the top panel were drawn in reverse,
i.e., with the position angle shifted by 180\arcdeg.
($b$)
Maps of the \Ht\ line ({\it heat scale}) and the SiO line ({\it contours})
toward the NGC 1333 IRAS 4 region.
The SiO map is the same as the one shown in Fig. 3$a$ of Paper I.
{\it Dashed curve}:
The parabolic fit in ($a$) drawn in the image coordinate.
{\it Large cross}:
\HCOp\ peak position of the dense core
obstructing the northeastern outflow (Choi 2005b).
{\it Filled circles}:
Millimeter continuum sources (Looney et al. 2000).}
\end{figure*}

%% file: fig4.tex
\begin{figure*}[t]
\centerline{\plotpanel{sio.h2-lowq.eps}{104mm}{1}\hspace*{5mm}%
            \plotpanel{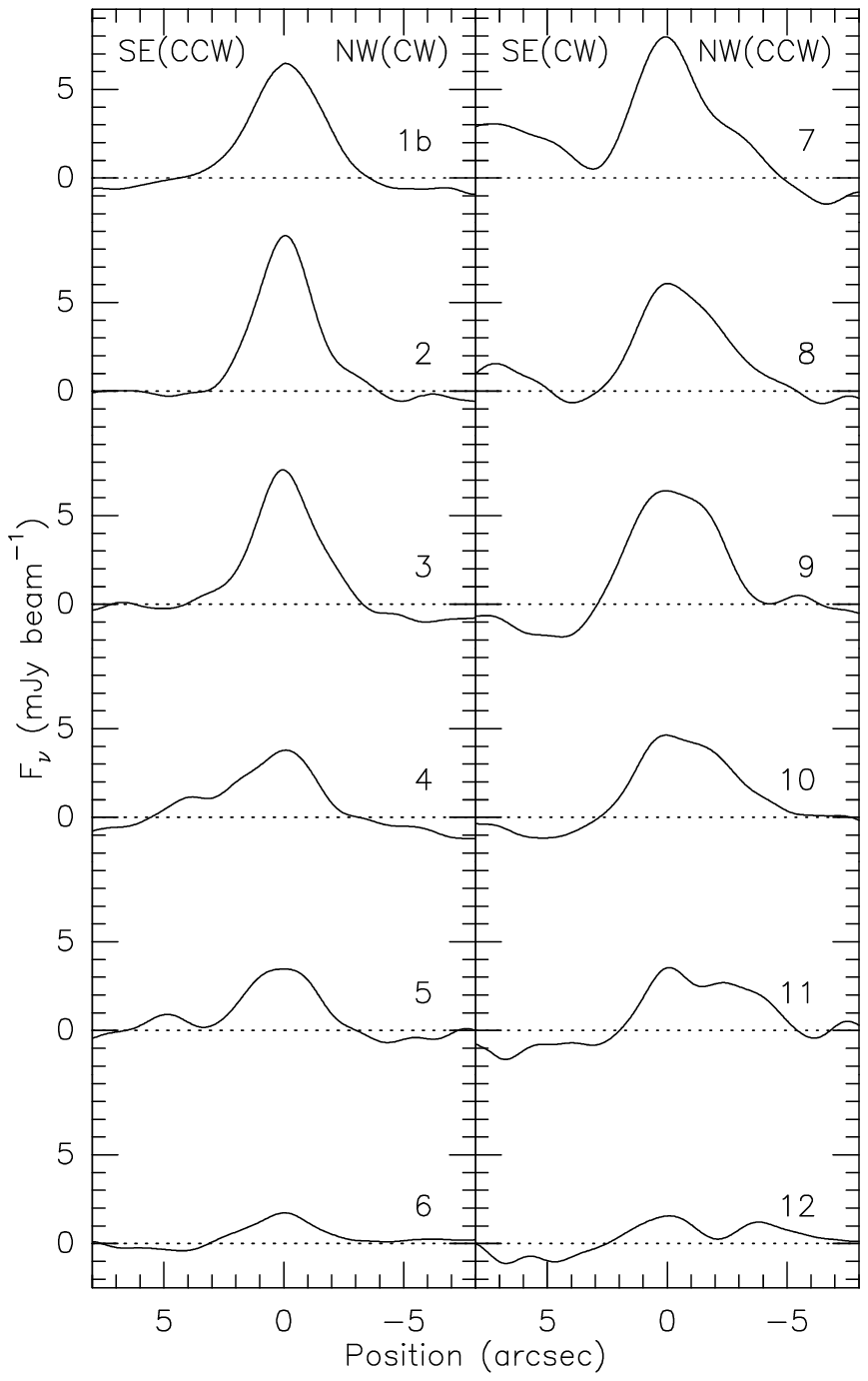}{76mm}{1}}
\caption{\small\baselineskip=0.825\baselineskip
($a$)
Maps of the SiO line toward the NGC 1333 IRAS 4 region.
See Figs. 2$a$ and 6 of Paper I for details.
Shown at the bottom left-hand corner is
the restoring beam: FWHM = 1\farcs96.
{\it Straight lines}:
Cuts for the intensity profiles.
They are perpendicular to the flow axis of the main bipolar outflow
(the solid straight lines in Fig. 2$a$ of Paper I).
{\it Crosses}:
Peak positions of the SiO emission.
Each peak of the main outflow is labeled
at the end of the corresponding cut.
{\it Heat scale}:
\Ht\ line image from the Subaru telescope (Fig. 2).
($b$)
Intensity profile of the SiO emission across the outflow lobe.
The narrow-line component in the southwestern region of the map
is not included (see Paper I).
The position axis is the angular distance along the cut
in arcseconds from each SiO outflow peak position.
The left-hand edge of each panel
corresponds to the southeastern end of the cut.
In the northeastern outflow lobe ({\it left panel}),
the clockwise side means the northwestern side.
It is the other way around in the southwestern outflow lobe
({\it right panel}).
The SiO peak number is labeled
on the right-hand side of each intensity profile.}
\end{figure*}

%% file: fig5.tex
\begin{figure}[t]
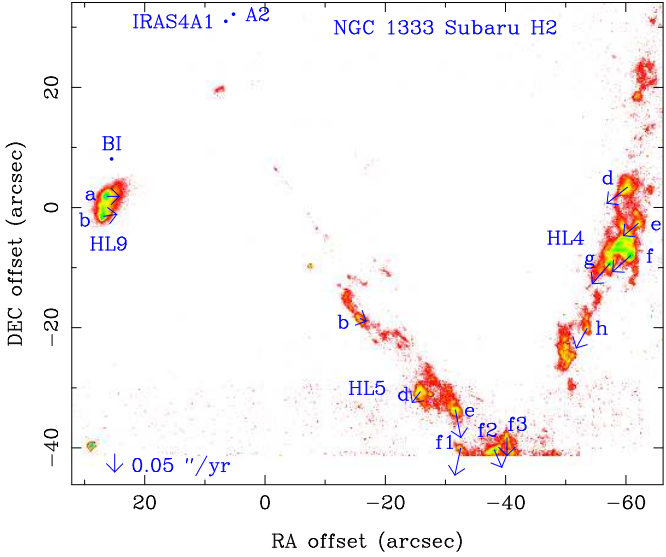

\ploteps{pmotion.eps}{88mm}
\caption{\small\baselineskip=0.825\baselineskip
Proper-motion vectors superposed on the Subaru \Ht\ image.
The length of arrows is proportional to the amount of proper motion,
and the size of arrow heads corresponds to the uncertainty.
The arrow at the bottom left-hand corner
shows the proper-motion scale.}
\end{figure}

%% file: fig6.tex
\begin{figure}[t]
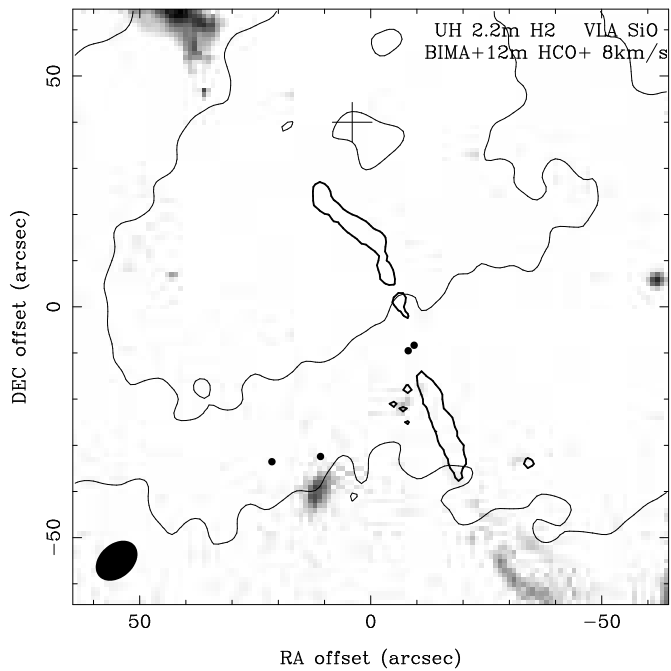

\ploteps{3maps.eps}{88mm}
\caption{\small\baselineskip=0.825\baselineskip
Maps of the \Ht\ line ({\it gray scale}),
the SiO line ({\it thick contours}, Paper I),
and the 8 \kms\ component of the \HCOp\ \JJ10\ line
({\it thin contours}, Choi 2005b).
The \HCOp\ and the SiO maps are not corrected for the primary beam response.
Shown at the bottom left-hand corner
is the synthesized beam of the \HCOp\ map.
The markers are the same as those in Fig. 3$b$.}
\end{figure}

%% file: tab1.tex
\begin{deluxetable}{p{22mm}cc}
\tabletypesize{\small}
\tablecaption{H$_2$ Emission Features}%
\tablewidth{0pt}
\tablehead{
& \multicolumn{2}{c}{\sc Peak Position} \\
\cline{2-3}
\colhead{\sc Source}
& \colhead{$\alpha_{\rm J2000.0}$} & \colhead{$\delta_{\rm J2000.0}$}}%
\startdata
CHH 1\dotfill  & 03 29 00.7 & 31 16 05 \\
CHH 2\dotfill  & 03 29 14.3 & 31 15 47 \\
CHH 3\dotfill  & 03 29 18.1 & 31 14 42 \\
CHH 4\dotfill  & 03 29 17.2 & 31 14 31 \\
CHH 5a\dotfill & 03 29 04.7 & 31 14 25 \\
CHH 5b\dotfill & 03 29 04.1 & 31 14 46 \\
CHH 5c\dotfill & 03 29 04.1 & 31 15 00 \\
CHH 6\dotfill  & 03 29 02.3 & 31 14 22 \\
CHH 7\dotfill  & 03 29 21.0 & 31 13 40 \\
CHH 8\dotfill  & 03 29 19.1 & 31 13 21 \\
CHH 9\dotfill  & 03 29 21.2 & 31 12 15 \\
CHH 10\dotfill & 03 29 18.3 & 31 12 10 \\
CHH 11\dotfill & 03 29 13.3 & 31 12 08 \\
CHH 12\dotfill & 03 28 59.6 & 31 12 05 \\
CHH 13\tablenotemark{a}\dotfill
               & 03 28 57.6 & 31 12 09 \\
CHH 14\dotfill & 03 29 16.4 & 31 09 51 \\
CHH 15\dotfill & 03 29 10.1 & 31 13 21 \\
CHH 16\dotfill & 03 29 08.7 & 31 13 07 \\
CHH 17\dotfill & 03 29 09.9 & 31 13 06 \\
CHH 18\dotfill & 03 29 09.3 & 31 12 56 \\
CHH 19\dotfill & 03 29 08.5 & 31 12 49 \\
\enddata\\
\tablecomments{Units of right ascension are hours, minutes, and seconds,
               and units of declination are degrees, arcminutes,
               and arcseconds.}%
\tablenotetext{a}{CHH 13 is at the edge of the field of view (Fig. 1).
                  The actual peak position of CHH 13 may be located
                  outside the region imaged.}%
\end{deluxetable}

%% file: ms.bbl
\begin{references}
\reference{} Aspin, C., Sandell, G., \& Russell, A. P. G. 1994,
             A\&AS, 106, 165
\reference{} Bally, J., Devine, D., \& Reipurth, B. 1996, ApJ, 473, L49
\reference{} Bate, M. R., Bonnell, I. A., Clarke, C. J., Lubow, S. H.,
             Ogilvie, G. I., Pringle J. E., \& Tout, C. A. 2000,
             MNRAS, 317, 773
\reference{} Blake, G. A., Sandell, G., van Dishoeck, E. F.,
             Groesbeck, T. D., Mundy, L. G., \& Aspin, C. 1995,
\reference{} Choi, M. 2001a, ApJ, 550, 817
\reference{} Choi, M. 2001b, ApJ, 553, 219
\reference{} Choi, M. 2005a, ApJ, 630, 976 (Paper I)
\reference{} Choi, M. 2005b, Journal of the Korean Physical Society, 47, 533
\reference{} Choi, M, Kamazaki, T., Tatematsu, K., \& Panis, J.-F. 2004,
             ApJ, 617, 1157
\reference{} Davis, C. J., Dent, W. R. F., Matthews, H. E., Coulson, I. M.,
             \& McCaughrean, M. J. 2000, MNRAS, 318, 952
\reference{} de Gouveia Dal Pino, E. M. 1999, ApJ, 526, 862
\reference{} de Zeeuw, P. T., Hoogerwerf, R., de Bruijne, J. H. J.,
             Brown, A. G. A., \& Blaauw, A. 1999, AJ, 117, 354
\reference{} Eisl{\"o}ffel, J., \& Mundt, R. 1997, AJ, 114, 280
\reference{} Fendt, C., \& Zinnecker, H. 1998, A\&A, 334, 750
\reference{} Girart, J. M., Crutcher, R. M., \& Rao, R. 1999, ApJ, 525, L109
\reference{} Hodapp, K. W., Bally, J., Eisl{\"o}ffel, J.,
             \& Davis, C. J. 2005, AJ, 129, 1580
\reference{} Hodapp, K.-W., et al. 1996, New Astronomy, 1, 177
\reference{} Hodapp, K.-W., \& Ladd, E. F. 1995, ApJ, 453, 715
\enlargethispage{-14\baselineskip}
\reference{} Lay, O. P., Carlstrom, J. E., \& Hills, R. E. 1995,
             ApJ, 452, L73
\reference{} Looney, L. W., Mundy, L. G., \& Welch, W. J. 2000, ApJ, 529, 477
\reference{} Machida, M. N., Matsumoto, T., Hanawa, T., \& Tomisaka, K. 2006,
             ApJ, in press
\reference{} Matsumoto, T., \& Tomisaka, K. 2004, ApJ, 616, 266
\reference{} Motohara, K., et al. 2002, PASJ, 54, 315
\reference{} Raga, A. C., \& Cant{\'o}, J. 1995, Rev. Mex. AA, 31, 51
\reference{} Raga, A. C., de Gouveia Dal Pino, E. M., Noriega-Crespo, A.,
             Mininni, P. D., \& Vel{\'a}zquez, P. F. 2002, A\&A, 392, 267
\reference{} Reipurth, B., Raga, A. C., \& Heathcote, S. 1996,
             A\&A, 311, 989
\reference{} Rodr\'{\i}guez, L. F., Anglada, G., \& Curiel, S. 1999,
             ApJS, 125, 427
\reference{} Sandell, G., Aspin, C., Duncan, W. D., Russell, A. P. G.,
             \& Robson, E. I. 1991, ApJ, 376, L17
\reference{} Sandell, G., \& Knee, L. B. G. 2001, ApJ, 546, L49
\reference{} Shepherd, D. S., Yu, K. C., Bally, J., \& Testi, L. 2000,
             ApJ, 535, 833
\reference{} Walawender, J., Bally, J., \& Reipurth, B. 2005, AJ, 129, 2308
\end{references}
